# The distribution of pairwise peculiar velocities in the nonlinear regime

Ravi K. Sheth
*Berkeley Astronomy Department, University of California, Berkeley, CA 94720, USA*

26 September 1995

**ABSTRACT**
The distribution of pairwise, relative peculiar velocities, $f(u; r)$, on small nonlinear scales, $r$, is derived from the Press–Schechter approach. This derivation assumes that Press–Schechter clumps are virialized and isothermal. The virialized assumption requires that the circular velocity, $V_c \propto M^{1/3}$, where $M$ denotes the mass of the clump. The isothermal assumption means that the circular velocity is independent of radius. Further, it is assumed that the velocity distribution within a clump is Maxwellian, that the pairwise relative velocity distribution is isotropic, and that on nonlinear scales clump-clump motions are unimportant when calculating the distribution of velocity differences.

Comparison with $N$-body simulations shows that, on small nonlinear scales, all these assumptions are accurate. For initially scale-invariant Gaussian density fields, the pairwise velocity distribution evolves in a self-similar manner. This is consistent with other analytic expectations, and with the distribution measured in relevant $N$-body simulations. For most power-spectra of interest, the resulting line of sight, pairwise, relative velocity distribution, $f(u_r)$, is well approximated by an exponential, rather than a Gaussian distribution.

This simple Press–Schechter model is also able to provide a natural explanation for the observed, non-Maxwellian shape of $f(v)$, the distribution of peculiar velocities.

## 1 INTRODUCTION

When quantifying the relation between real and redshift space galaxy catalogs, the distribution function of pairwise peculiar velocities is of fundamental importance. In the linear regime the effects of redshift induced distortions can be calculated explicitly (e.g. Peebles 1980). On large, linear scales, the correlation function in redshift space is closely related to the correlation function in real space (Kaiser 1986; Hamilton 1992). In the nonlinear regime, also, the correlation functions in real and redshift space may be easily related, provided that on small, nonlinear scales the clustering is stable so that, on average, the pairwise motions (on these small scales) exactly cancel the Hubble expansion. However, to relate the two correlation functions requires knowledge of the distribution of pairwise relative peculiar velocities (Peebles 1980, §76). It is known from the linear theory that if the initial density field is Gaussian, then the pairwise relative velocity distribution that obtains on large, linear scales must also be Gaussian (e.g. Fisher 1995). In contrast, on smaller nonlinear scales, observations (Peebles 1976) and $N$-body simulations (Efstathiou et al. 1988; Fisher et al. 1994) suggest that the pairwise relative velocity distribution, when viewed along the line of sight, is better approximated by an exponential, than by a Gaussian distribution. This paper shows that the accuracy of the exponential distribution on nonlinear scales can be easily explained by the Press–Schechter theory of nonlinear clustering.

Section 2 shows how to calculate the distribution of pairwise, relative peculiar velocities. It restricts attention to the highly nonlinear regime, since when pair separations are small the distribution of pairwise velocities should be insensitive to clump-clump motions. Essentially, this is because particles that are separated by small distances are most likely to be in the same clump, so the motion of the clump as a whole is irrelevant when computing the pairwise relative velocity. The calculation assumes that Press–Schechter clumps are virialized and isothermal, and that the velocity distribution of particles within each clump is Maxwellian. For initially scale-invariant power-spectra, the distribution of pairwise relative peculiar velocities that results evolves self-similarly, in accordance with the prescription discussed by Davis & Peebles (1977), and tested by Efstathiou et al. (1988). For all power-spectra of current interest, the distri-bution of line of sight, pairwise, peculiar velocities is well fit by an exponential, rather than by a Gaussian distribution.

Section 3 shows how to use the model developed in Section 2 to calculate the peculiar velocity distribution function (as opposed to the distribution of velocity differences) using the Press–Schechter approach. This distribution is the analogue of the Maxwell-Boltzmann distribution that describes an ideal gas. Unlike the pairwise, relative peculiar velocity distribution computed in the previous section, this distribution can only be derived assuming some model for the motions of clumps towards each other. Section 3 presents a simple, *ad hoc* model for this clump-clump velocity distribution. In this respect, the results are Section 3 are less general than those of Section 2, since they dependend on assumptions about the motions of clumps towards each other. The peculiar velocity distribution function that results is well fit by a functional form, used previously by Saslaw et al. (1990), that fits $N$-body simulations of clustering from Poisson initial conditions well. It is interesting that this distribution also appears to be consistent with the measured peculiar motions of galaxies (Raychaudhury & Saslaw 1995) that are available at present.

Section 4 summarizes the results and discusses the relation between the assumption that clustering on nonlinear scales is statistically stable (i.e., that, on average, the pairwise velocities cancel the Hubble flow exactly), and the Press–Schechter model that is developed and extended in this paper.

## 2 THE DISTRIBUTION OF PAIRWISE, RELATIVE PECULIAR VELOCITIES

This section develops a model for the distribution of pairwise, relative peculiar velocities for pairs that are separated by small distances.

### 2.1 The model

Let $f(\boldsymbol{v})\,d\boldsymbol{v}$ denote the probability that the peculiar velocity of a particle is in the range $d\boldsymbol{v}$ about $\boldsymbol{v}$. Also, write the peculiar velocity of the $i$th particle as the sum of two terms:

$$\boldsymbol{v}_i = \boldsymbol{v}_{\mathrm{p}_i} + \boldsymbol{v}_{\mathrm{c}_i}, \tag{1}$$

where the first term on the right describes the motion of the particle with respect to the center of mass of the clump of which it is a member, and the second term describes the motion of the clump as a whole. Then the pairwise velocity distribution for pairs separated by the vector $\boldsymbol{r}$, $f(\boldsymbol{u};\boldsymbol{r}) = f(\boldsymbol{v}_1 - \boldsymbol{v}_2;\boldsymbol{r})$, requires knowledge of the joint velocity distribution, $f(\boldsymbol{v}_1, \boldsymbol{v}_2)$, where particles 1 and 2 are separated by $\boldsymbol{r}$. This with equation (1) shows that, in general, to compute $f(\boldsymbol{u};\boldsymbol{r})$ requires knowledge of the pairwise velocity distribution of clumps. Usually, these clump-clump velocities are not known. However, if we restrict attention to separations that are sufficiently small that both members of most pairs are likely to be in the same clump, then $f(\boldsymbol{u};\boldsymbol{r})$ can be obtained without any knowledge of the motions of clumps, since then $\boldsymbol{u}_{ij} = \boldsymbol{v}_{\mathrm{p}_i} - \boldsymbol{v}_{\mathrm{p}_j}$ for most pairs. So, provided we restrict attention to separations that are small compared to the dimensions of a typical clump, the distribution of pairwise velocities can be obtained without any knowledge of the motions of clumps. Clearly, however, the distribution of pairwise velocities, even on these small scales, depends on the distribution of velocities within a given clump, and on the distribution of clump sizes. So, for small separations, the pairwise velocity distribution can be written as

$$f(\boldsymbol{u};\boldsymbol{r}) = \int p(\boldsymbol{u}|M;\boldsymbol{r})\,p(M;\boldsymbol{r})\,\mathrm{d}M, \tag{2}$$

where $p(\boldsymbol{u}|M;\boldsymbol{r})$ is the probability that the pairwise velocity difference for particles separated by $\boldsymbol{r}$ in a clump of mass $M$ is $\boldsymbol{u}$, the second term gives the probability that the pair is within a clump of mass $M$, and the integral is over all clumps that are larger than $|\boldsymbol{r}|$. In particular,

$$p(M;\boldsymbol{r})\,\mathrm{d}M = \frac{N_{\mathrm{pairs}}(M;\boldsymbol{r})\,\mathrm{d}M}{\int N_{\mathrm{pairs}}(M;\boldsymbol{r})\,\mathrm{d}M}, \tag{3}$$

where the denominator denotes the total number of pairs that have the separation vector $\boldsymbol{r}$. For small values of $|\boldsymbol{r}|$, this total number of pairs may be approximated by assuming that both members of each pair are in the same clump, so that the integral is over all clumps that are larger than $|\boldsymbol{r}|$. In what follows, we will assume that the distribution of $\boldsymbol{u} = \boldsymbol{v}_1 - \boldsymbol{v}_2$ is isotropic. On the small scales of interest, this is likely to be a good approximation (e.g. Efstathiou et al. 1988). This means that $p(\boldsymbol{u}|M;\boldsymbol{r})$ and $p(M;\boldsymbol{r})$ may be replaced with $p(u|M;r)$ and $p(M;r)$, respectively. With this assumption of isotropy, $f(u;r)$ will be computed by using the Press–Schechter description of nonlinear gravitational clustering (Press & Schechter 1974; Bond et al. 1991) to construct models for both $p(u|M;r)$ and $p(M;r)$.

In the Press–Schechter approach, the number density of clumps of a given mass can be computed directly from the statistics of the initial density field. These Press–Schechter clumps are assumed to virialize when they are $\sim 178$ times more dense than the background density. This fact will be useful when computing the number of pairs as a function of separation, $p(M;r)$. A virialized clump of mass $M$ may be assigned a circular velocity, $V_c$, in accordance with the prescription suggested by Narayan & White (1988), and used later by White & Frenk (1991). Namely, the assumption that $V_c^2 = GM/r_{\mathrm{vir}}$, where $V_c$ defines a circular velocity that can be associated with the clump of mass $M$, and $r_{\mathrm{vir}}$ denotes the radius of the clump (i.e., $\rho_{\mathrm{vir}} = 178\rho_{\mathrm{b}} = 3M/4\pi r_{\mathrm{vir}}^3$), provides a relation between the circular velocity $V_c$ and the mass: $V_c \propto M^{1/3}$. Assuming that these virialized clumps are also isothermal means that $V_c$ is approximately independent of position within the clump. $N$-body simulations show that the $V_c \propto M^{1/3}$ scaling is quite accurate (e.g., Bond & Myers, in preparation; but see discussion at the end of this paper). The simulations of Efstathiou et al. (1988) show that, within a given clump, $V_c$ is, indeed, approximately independent of radius.

For an isothermal sphere $V_c^2$ is two-thirds of the three dimensional velocity dispersion (cf. Section 4.4.3(b) in Binney & Tremaine 1987; Appendix B in Lacey & Cole 1993). Therefore, in this model, particles in a clump of mass $M$ are assigned velocities from a Maxwellian distribution that has a three-dimensional mean square velocity given by $3V_c^2(M)/2$. The choice of a Maxwellian for the velocity distribution within each clump is rather *ad hoc*; although it is suggested by the virial assumption, it is not required. It may be that one of the self-consistent velocity distributions derived by Evans (1994) for density distributions that vary as power-laws of the radius, is more appropriate. The Maxwellian is chosen here simply to provide some insight into the shape of the velocity distribution when it is averaged over all Press–Schechter clumps.

Thus, in this model, the distribution of pairwise velocities on small scales is given by calculating the distribution of the vector $\boldsymbol{u} = \boldsymbol{v}_1 - \boldsymbol{v}_2$, where $\boldsymbol{v}_1$ and $\boldsymbol{v}_2$ are Maxwellian vectors. For small separations, it is very likely that both members of a given pair are members of the same clump. So, the assumption that the clump is isothermal means that the Maxwellian distribution that describes $\boldsymbol{v}_1$ also describes the distribution of $\boldsymbol{v}_2$. Therefore, the distribution of $\boldsymbol{u} = \boldsymbol{v}_1 - \boldsymbol{v}_2$ is also Maxwellian, since calculating the distribution of $\boldsymbol{v}_1 - \boldsymbol{v}_2$ reduces to computing the distribution of $(v_{x1} - v_{x2}, v_{y1} - v_{y2}, v_{z1} - v_{z2})$, where the $v_{ij}$s are all Gaussian distributed. Since it is obtained from the difference of two Gaussian random variables that have the same mean and variance as each other, the mean square of this pairwise Maxwellian velocity distribution is $3V_c^2(M)$ (this is essentially Problem 7-3 on p.485 in Binney & Tremaine 1987). This follows because the three dimensional dispersion of the Maxwellian that describes the motions of the particles themselves is $3V_c^2(M)/2$, and we are noting explicitly that the pair belongs to a clump of mass $M$. Thus, the assumption that the Press–Schechter clumps are virialized isothermal spheres implies that the distribution $p(\boldsymbol{u}|M;r)$ of equation (2) is Maxwellian, with a dispersion that is related to the mass of the clump.

The distribution of pairs, $p(M;r)$, can be obtained from the assumption that all Press–Schechter clumps are truncated, singular isothermal spheres having the same density ($\sim$178 times the background density). The Appendix shows that if the distribution of particles is isotropic, then the number of pairs separated by a distance $r$ that are in a clump of mass $M$ and radius $R$ scales as $N_{\rm pairs}(M;r) \propto M^2/R^2$, at least to first order. Since all clumps have the same density, $R^3 \propto M$, so that $N_{\rm pairs}(M;r) \propto M^{4/3}$. Thus,

$$p(M;r)\,{\rm d}M \approx \frac{M^{4/3}\, n(M)\,{\rm d}M}{\int_{M_{\rm min}}^{\infty} M^{4/3}\, n(M)\,{\rm d}M}, \qquad (4)$$

where $n(M)$ is the number density of clumps of mass $M$. Note that, for consistency, the integral in the denominator is only over those clumps whose diameters, $2R$, are larger than the pair separation $r$. Since all clumps have the same density, this implies that $M_{\rm min} \propto R_{\rm min}^3$, with $R_{\rm min} = r/2$. Strictly speaking, equation (4) for the pair distribution is correct only for pair separations that are small compared to the clump radius. The exact expression, given in the Appendix, is complicated. For the sake of simplicity (and because it will turn out that the results are not very sensitive to this approximation), we will continue to use the simpler equation (4) in the remainder of this section.

Since $n(M)$ is given by the Press–Schechter distribution, $f(u;r)$ can now be obtained by computing the integral in equation (2). The assumption of isotropy allows us to obtain the line of sight distribution from equation (2). Since the velocity dependence is contained entirely in the Maxwellian $p(\boldsymbol{u}|M;r)$ term, it is sufficient to integrate this Maxwellian over all directions perpendicular to the line of sight. The Maxwellian reduces to a Gaussian, so that the line of sight distribution is a Gaussian (with dispersion $\sigma^2 = 2\sigma_{3D}^2 = 3V_c^2 \propto 3M^{2/3}$) convolved with the $p(M;r)$ distribution of equation (4):

$$f(u;r) = \int_{M_{\rm min}}^{\infty} \frac{e^{-\frac{3}{2}\frac{u^2}{\sigma^2}}}{\sqrt{2\pi\sigma^2/3}}\, p(M;r)\,{\rm d}M, \qquad (5)$$

where $u$ now denotes the line of sight, rather than the three dimensional velocity. Once $n(M)$ in $p(M;r)$ has been specified, $f(u;r)$ may be obtained by integrating equation (5). Equation (5) is the main, general result of this paper. It shows that the line of sight pairwise velocity distribution is essentially a weighted sum over Gaussians having a range of dispersions, where the weighting factor for each Gaussian is related to the Press–Schechter multiplicity function, and to the distribution of particles within a Press–Schechter clump.

### 2.2 Relation to BBGKY scaling solution

This subsection considers a useful, specific example of the model derived above; the evolution of the pairwise velocity distribution as an initially scale free density field clusters gravitationally.

For an initially scale free field, Davis & Peebles (1977) argue that the BBGKY equations can be transformed into a series of equations for the moments of the pairwise velocity distribution. They show that the $N$-point correlation functions admit a similarity transformation of the form

$$s = \frac{x}{t^A}, \qquad {\rm where} \qquad A = \frac{4}{3(3+n)}, \qquad (6)$$

where $n$ is the slope of the power-spectrum of the initially scale invariant Gaussian density field [i.e., the initial power spectrum is $P(k) \propto k^n$], $t$ denotes cosmic time, $a$ denotes the expansion factor at that time ($a \propto t^{2/3}$ in a Universe that has critical density), and $x$ is a comoving distance. They also argue that if equation (6) is satisfied, then the BBGKY hierarchy requires that the relative peculiar velocities be rescaled by the relation

$$\hat{u} = \frac{u}{a\,t^{A-1}} \qquad (7)$$

(e.g. Efstathiou et al. 1988). We can compare this scaling solution to the model described in the previous subsection.

Recall that $GM/r_{\rm vir} = V_c^2 = GM/(3M/4\pi\Delta_{\rm nl}\rho_{\rm b})^{1/3}$, where $\Delta_{\rm nl} \sim 178$ and $\rho_{\rm b} \propto a^{-3}$ is the background density, and that $V_c^2 = \sigma^2/3$, where $\sigma^2$ is the three dimensional velocity dispersion of the pairwise velocity distribution. In a Universe with critical density (i.e. $3H^2 = 8\pi G\rho_{\rm b}$), these relations imply that

$$M = \left(\frac{3}{4\pi G \Delta_{\rm nl}\rho_{\rm b}}\right)^{1/2} \frac{V_c^3}{G} = \frac{\sqrt{2}}{3^{3/2} GH\Delta_{\rm nl}^{1/2}}\,\sigma^3 \equiv \alpha\,\sigma^3. \quad (8)$$

Equation (8) defines $\alpha$, and $\alpha \propto a^{3/2}$ since $\rho_{\rm b} \propto a^{-3}$. This, in equation (5), gives

$$f(u;r) = \frac{\int_{M_{\rm min}}^{\infty} \sqrt{\frac{3}{2\pi\sigma^2}}\, e^{-\frac{3}{2}\frac{u^2}{\sigma^2}}\, M^{4/3}\, n(M)\,{\rm d}M}{\int_{M_{\rm min}}^{\infty} M^{4/3}\, n(M)\,{\rm d}M} \qquad (9)$$

where

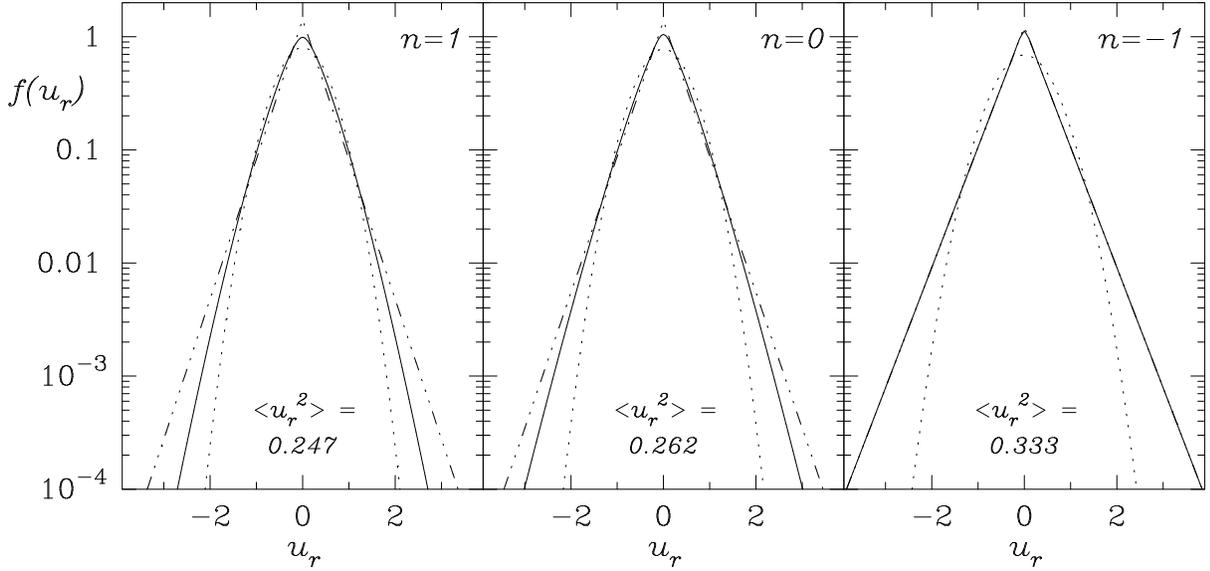

**Figure 1.** Examples of the distribution function of pairwise relative peculiar velocities. In each panel, solid line shows equation (15). The corresponding value of the mean square pairwise velocity (eq. 16) is also shown. Dot-dashed and dotted lines show exponential and Gaussian distributions that have the same dispersion as the solid curves.

$$n(M)\,\mathrm{d}M = \frac{1}{\sqrt{\pi}}\left(\frac{M}{M_*}\right)^{\frac{n+3}{6}} e^{-\left(\frac{M}{M_*}\right)^{\frac{n+3}{3}}} \left(\frac{n+3}{3}\right) \frac{\mathrm{d}M}{M^2} \quad (10)$$

is the Press–Schechter distribution for an initially scale free field,

$$M_* = C_n\, a^{6/(3+n)}, \quad (11)$$

and $C_n$ is a constant that depends on the slope $n$ and amplitude of the initial power spectrum (e.g. Efstathiou et al. 1988). Now define

$$\hat{u} \equiv (\alpha/M_*)^{1/3}\, u \propto \frac{a^{1/2}}{a^{2/(3+n)}}\, u \propto \frac{a^{1/2}}{t^{4/3(3+n)}}\, u \propto \frac{u}{a t^{A-1}}, \quad (12)$$

where $A$ was defined in equation (6). So $\hat{u}$ scales similarly to the Davis & Peebles solution. Also, set $\nu \equiv M/M_*$. Then

$$f(\hat{u};\hat{r}) = f(u;r)\frac{\mathrm{d}u}{\mathrm{d}\hat{u}} = f(u;r)\left(\frac{M_*}{\alpha}\right)^{1/3}$$

$$= \frac{\int_{\nu_{\min}}^\infty \sqrt{\frac{3}{2\pi}} e^{-\frac{3}{2}\frac{\hat{u}^2}{\nu^{2/3}}}\, \nu^{\frac{n+3}{6}}\, e^{-\nu^{\frac{n+3}{3}}} \left(\frac{n+3}{3}\right) \frac{\mathrm{d}\nu}{\nu}}{\gamma\left(\frac{1}{2} + \frac{1}{n+3}, \nu_{\min}^{\frac{n+3}{3}}\right)}, \quad (13)$$

where the denominator in the final expression is the complimentary incomplete Gamma function. This shows explicitly that if $\nu_{\min}$ is constant, then $f(\hat{u};\hat{r})$ evolves self-similarly. However, recall that $M_{\min}$ is related to $R_{\min}$ and $R_{\min} = r/2$, where $r$ is the separation scale. This means that

$$\nu_{\min} = \frac{M_{\min}}{M_*} = \frac{4\pi\Delta_{\rm nl}\rho_o}{3 C_n}\left(\frac{R_{\min}/a}{t^{4/3(3+n)}}\right)^3. \quad (14)$$

Therefore, requiring that $\nu_{\min}$ remain constant implies that $R_{\min}/a$, the comoving cutoff distance, must scale as $t^{4/3(3+n)}$. Comparison with equation (6) shows that this is just what is required by the Davis–Peebles solution. Thus, this model for the pairwise velocity distribution satisfies the Davis–Peebles scaling solution of the BBGKY hierarchy.

Having shown that this Press–Schechter model for the pairwise velocity distribution on small scales has the correct scaling, we now consider the shape of this distribution. First, consider the limit of very small pair separations (so that $M_{\min} \to 0$). The pairwise velocity distribution becomes

$$f(\hat{u}) = \frac{\int_0^\infty \sqrt{\frac{3}{2\pi}} e^{-\frac{3}{2}\frac{\hat{u}^2}{\nu^{2/3}}}\, \nu^{\frac{n+3}{6}}\, e^{-\nu^{\frac{n+3}{3}}} \left(\frac{n+3}{3}\right) \frac{\mathrm{d}\nu}{\nu}}{\int_0^\infty \nu^{\frac{n+3}{6}+\frac{1}{3}}\, e^{-\nu^{\frac{n+3}{3}}} \left(\frac{n+3}{3}\right) \frac{\mathrm{d}\nu}{\nu}}$$

$$= \frac{\int_0^\infty \sqrt{\frac{3}{2\pi}} e^{-\frac{3}{2}\frac{\hat{u}^2}{x^2}}\, x^{\frac{n+3}{2}}\, e^{-x^{(n+3)}} \left(\frac{n+3}{2}\right) \frac{\mathrm{d}x^2}{x^2}}{\Gamma\left(\frac{1}{2}+\frac{1}{n+3}\right)}, \quad (15)$$

where $x^2 = \nu^{2/3} = (M/M_*)^{2/3}$. In this limit, the pairwise dispersion is

$$\langle \hat{u}^2 \rangle = \frac{1}{3}\frac{\Gamma\left(\frac{1}{2}+\frac{3}{n+3}\right)}{\Gamma\left(\frac{1}{2}+\frac{1}{n+3}\right)}. \quad (16)$$

Equation (16) shows that for $n = 1, 0,$ and $-1$, the scaled rms line of sight velocities are $\sqrt{\langle \hat{u}^2\rangle} = \sqrt{0.2466}, \sqrt{0.2617}$, and $\sqrt{1/3}$, respectively.

When $n = -1$, then equation (15) can be calculated analytically:

$$f(\hat{u}) = \int_0^\infty \sqrt{\frac{3}{2\pi}}\, e^{-\frac{3}{2}\frac{\hat{u}^2}{x^2}}\, x\, e^{-x}\, \frac{\mathrm{d}x^2}{x^2}$$

$$= \sqrt{\frac{3}{2}}\, \exp\left(-2\sqrt{3/2}\,\hat{u}\right). \quad (17)$$

The final expression follows from equation (3.325) in Gradshteyn & Ryzhik (1994). Equation (17) shows that when the initial fluctuation spectrum is scale free with slope $n = -1$, then in the limit of vanishingly small pair separations, the probability that the pairwise velocity is in the range $du$ about $u$ is given by an exponential distribution.

Moreover, consider those initial conditions that have Press–Schechter $n(M)$ distributions that are not vastly different from the $n = -1$ shape. In the limit of very small pair separations, the exponential will provide a good approximation to $f(\hat{u})$ for these other initial conditions also. To illustrate this, Fig. 1 shows $f(\hat{u})$ distributions for $n = 1, 0$ and $-1$, in the limit where $\nu_{\min} = 0$. The solid line in each panel shows equation (15). The corresponding value of the mean square pairwise velocity (eq. 16) is also shown. Dot-dashed and dotted lines show exponential and Gaussian distributions that have the same dispersion as the solid curves. In all three panels, the solid curves (eq. 15) are well approximated by the dot-dashed curves (exponential distributions).

When the cutoff, $\nu_{\min}$, is not vanishingly small, then the shape of $f(\hat{u}; \hat{r})$ will depart from the solid curves shown in the figure. The extent of these departures may be estimated as follows. A given value of $\nu_{\min} = M_{\min}/M_*$ means that clumps less massive than $M_{\min}/M_* = (R_{\min}/r_*)^3$ do not contribute to $f(\hat{u}; \hat{r})$. Now, smaller clumps have smaller circular velocities, so they contribute to $f(\hat{u}; \hat{r})$ primarily when $|\hat{u}|$ is small. Therefore the existence of a nonvanishing $\nu_{\min}$ has two related effects. The first is that, for a given $\nu_{\min}$, the departures from the $\nu_{\min} = 0$ shape will be maximal near the small $|\hat{u}|$ peak of the distribution. This implies the second effect: As $\nu_{\min}$ increases, the dispersion will also increase.

However, this model for $f(\hat{u}; \hat{r})$ (eq. 9) only applies for small values of $\nu_{\min}$. To see this, recall that $r_*$ is the scale on which the mass variance in spherical volumes is $\approx \delta_c^2/2 \approx (1.68/a)^2/2$ (e.g., Efstathiou et al. 1988). So, $r_*$ should be slightly larger than $r_0$, the correlation length, which is approximately the average Press–Schechter clump size (Peebles 1980). Since this model is only accurate on scales that are smaller than the average clump size, $\nu_{\min} = (R_{\min}/r_*)^3 \approx (R_{\min}/r_0)^3$ should always be less than, say, $(1/2)^3 \sim 0.1$. For these small values of $\nu_{\min}$, the tails of $f(\hat{u}; \hat{r})$ should remain unchanged from their $\nu_{\min} = 0$ values. To see the effects of $\nu_{\min}$ directly, equation (9) must be integrated numerically. Fig. 2 in the next subsection shows the result of integrating equation (9) for a range of values of $n$, and for representative values of the cutoff.

### 2.3 Comparison with $N$-body simulations

The Press–Schechter model for $f(\hat{u}; \hat{r})$ developed in the previous subsection may be compared with that measured in the $N$-body simulations of gravitational clustering from scale free initial conditions studied by Efstathiou et al. (1988). Efstathiou et al. define scaled distances and velocities using the transformations given in equations (6) and (7). They define $s_0$ as that value of the similarity parameter, $s$, at which the correlation function is unity, and present most of their velocity distribution results in terms of the variable $\hat{u}_{\rm EFWD}/s_0$, scaled to $a = 1$. They find that the scaled velocity distributions that they measure in their simulations have roughly exponential tails for the values of $n$ they consider. However, the peaks of the distributions are more rounded than an exponential (their Fig. 6).

Since the model developed in the previous subsection also produces exponential distributions (Fig. 1), we must check that it also gives the measured dispersions. To do so, we must relate the scaled velocity, $\hat{u}$, defined in equation (12) to $\hat{u}_{\rm EFWD}$, the scaled variable used by Efstathiou et al. Using equations (8), (11), (12), and the fact that $a_0 = 1$ and $C_{\rm n} =$ 0.8, 0.71, and 0.53 for $n = 1, 0$, and $-1$, respectively (their eq. 15), yields

$$\hat{u} \equiv \left(\frac{\alpha}{M_*}\right)^{\frac{1}{3}} u = \sqrt{\frac{1}{3}} \left(\frac{2}{\Delta_{\rm nl}}\right)^{\frac{1}{6}} \left(\frac{8\pi}{3C_{\rm n}}\right)^{\frac{1}{3}} \frac{32}{2\pi} \frac{3 t_0^A}{2} \frac{u}{a t^{A-1}}. \quad (18)$$

Since $\Delta_{\rm nl} = 178$, this means that

$$\frac{\hat{u}_{\rm EFWD}}{s_0} = \frac{0.236\, C_{\rm n}^{1/3}}{s_0 t_0^A}\, \hat{u}. \quad (19)$$

Their Fig. 4 shows that $s_0 t_0^A/L \approx 0.048/2\pi, 0.054/2\pi$ and $0.065/2\pi$, for $n = 1, 0$, and $-1$, respectively. These values, with their Fig. 6, suggest that $t_0 = 2/3$. In the limit of very small pair separations we can use equation (16) to estimate the variance of the model distributions:

$$\frac{\sqrt{\langle \hat{u}_{\rm EFWD}^2 \rangle}}{s_0} = \frac{0.236\, C_{\rm n}^{1/3}}{s_0 t_0^A} \sqrt{\frac{1}{3} \frac{\Gamma\left(\frac{1}{2} + \frac{3}{n+3}\right)}{\Gamma\left(\frac{1}{2} + \frac{1}{n+3}\right)}}. \quad (20)$$

Equation (20) implies that the scaled one dimensional rms pairwise dispersion in the simulations should take the values $\sqrt{\langle \hat{u}^2 \rangle}/s_0 = 2.2, 2.0$, and $1.7$, for $n = 1, 0$, and $-1$, respectively. These values differ slightly from those measured in the simulations, where the corresponding values are 2, 1.8, and 1.6 (the small $s$ limit in Fig. 5 of Efstathiou et al. 1988). Some of this discrepancy may be due to the fact that in the simulations, virialized clumps collapse by a factor of 1.8 rather than the virial factor two (Hamilton et al. 1991, who suggest that some of this may be due to force softening effects), so that $\Delta_{\rm nl}^{1/3}$ should be replaced by $0.9\Delta_{\rm nl}^{1/3}$. However, some of the discrepancy may also be due to the fact that not all clumps in the simulations are truncated singular isothermal spheres. Whatever the reason, these values for the dispersion are close enough to the $N$-body results to consider the model further.

Having shown that the Press–Schechter model has the correct scaling and approximately the right pairwise velocity dispersion, we now examine the shape of the distribution for nonzero values of the cutoff. Recall that this shape is given by equation (13) which must be integrated numerically. For ease of comparison with the $N$-body simulations, a representative value of the cutoff, $\nu_{\min}$, is chosen as follows. In their Fig. 4, Efstathiou et al. show the correlation functions measured in their simulations, and indicate the scales $x_{100}$ and $x_0$ at which $\xi = 100$ and $\xi = 1$. This allows us to estimate $x_{10}$. Their figure suggests that $\log_{10}(x_0/x_{10}) \approx 0.3, 0.38$ and $0.45$ for $n = 1, 0$, and $-1$, respectively. The last paragraph of the previous subsection argued that $\nu_{\min} \sim (R_{\min}/r_0)^3 \sim (x/x_0)^3$. So, setting $\nu_{\min} \approx (x_{10}/x_0)^3$ gives an indication of the shape of the velocity distribution for particles separated by the scale corresponding to $\xi \sim 10$.

Fig. 2 shows $f(\hat{u}_{\rm EFWD}/s_0)$ as a function of $\hat{u}_{\rm EFWD}/\hat{v}_{\rm H}$, when $n = 1, 0$, and $-1$ for this choice of $\nu_{\min}$. Velocities are in units of the Hubble velocity across the Efstathiou et al. (1988) computational volume, and have been rescaled using equation (19) to $a = 1$, for ease of comparison with their Fig. 6. (On the scale corresponding to $s$, the scaled value of the Hubble velocity is $\hat{v}_{\rm H} = 2s/3$.) Each panel shows the value of $n$ and the corresponding rescaled value of the mean square pairwise velocity. Solid lines show the distribution that is obtained by integrating equation (13) with $\nu_{\rm min} = 0.126, 0.075$, and $0.045$ for $n = 1, 0$, and $-1$, respectively. Dot-dashed lines show exponential distributions with the same dispersion as the solid curves, had the solid curves been computed with $\nu_{\rm min} = 0$. Comparison of the shapes of these curves with those in Fig. 1 shows that the effect of the cutoff is most pronounced near the peak of the distribution; the tails are hardly affected. This is consistent with the discussion at the end of the previous subsection. Dotted lines show Gaussian distributions (corresponding to Maxwellians in three-dimensions) with the same dispersions as the solid curves.

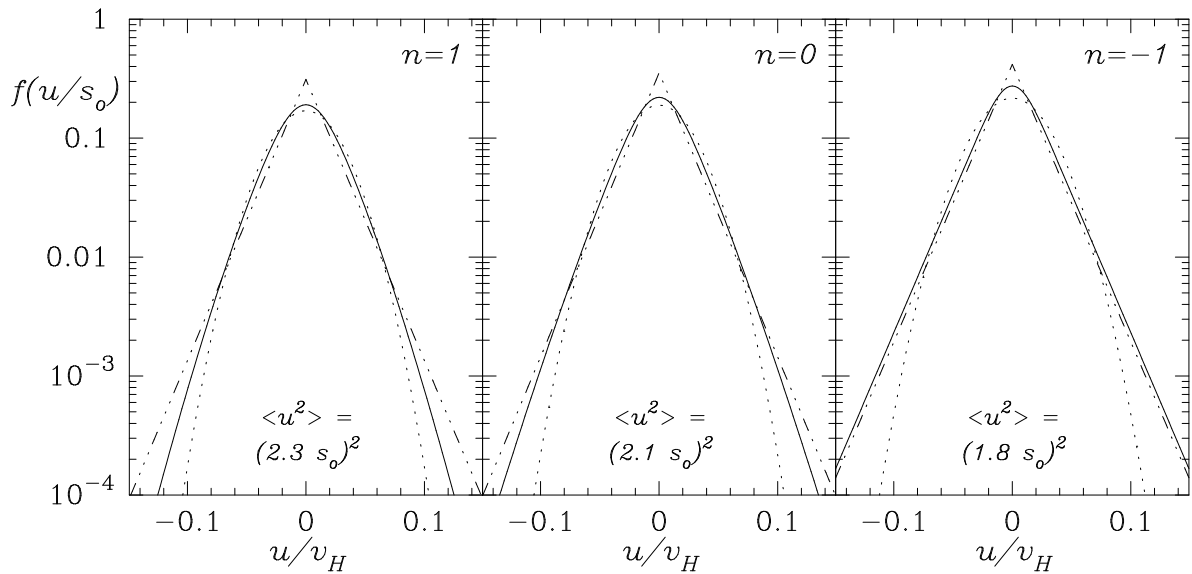

**Figure 2.** Examples of the distribution function of pairwise relative peculiar velocities on scales where $\xi \approx 10$. Velocities are in units that simplify comparison with the scaled $f(\hat{u}_{\rm EFWD}/s_0)$ distribution measured in $N$-body simulations (Fig. 6 of Efstathiou et al. 1988). In each panel, solid line shows equation (13), scaled to the $N$-body variables as described in the text (eq. 19). The corresponding scaled value of the mean square pairwise velocity is also shown. Dotted lines show Gaussian distributions that have the same dispersions as the solid curves. Dot-dashed lines show exponential distributions with dispersions given by equation (20); for these values of $n$, these dispersions are slightly less than those of the solid curves.

To conclude this subsection, we note that the solid curve in the leftmost panel of Fig. 2 ($n = 1$, on scale corresponding to $\xi \approx 10$) resembles the leftmost panel of Fig. 6 of Efstathiou et al. closely. Some of the small differences between the two figures may be due to the fact that whereas Efstathiou et al. show the $n = 1$ distribution for $\xi \approx 10$, the scale for Fig. 2 is only approximately where $\xi \approx 10$. Moreover, recall that Fig. 2 was constructed using an approximate expression for $p(M;r)$ (cf. eq. 4). Using the exact expression (given in the Appendix) makes the two figures even more alike. In any case, the figures are similar enough that we conclude that this Press–Schechter model for the pairwise velocity distribution is in good agreement with the $N$-body simulations.

### 2.4 The effects of discreteness

There is one complication when comparing these Press–Schechter models of clustering from initially Gaussian density fields with the $N$-body simulations. It arises from the fact that, in the Press–Schechter description, the probability of forming arbitrarily small clumps is not negligible. Since the velocity dispersion within a clump is related to its mass, this means that the probability of having arbitrarily small velocities is also not negligible. The $N$-body simulations, on the other hand, only have a limited dynamic range; they cannot resolve clumps that are smaller than some minimum size. In terms of the pairwise velocity distributions, this means that the simulations are only sensitive to clumps that have velocity dispersions greater than some minimum value. When comparing this Press–Schechter model with simulations, it is not clear what effect this discreteness will have when choosing the lower bound on the integration variable in equation (13).

The importance of discreteness in the Efstathiou et al. (1988) simulations can be estimated as follows. Assume that the particles in the simulations constitute a distribution of points that has the same correlation function as the continuous density field that the particles are being used to simulate. One way to do this is to assume that the particles in an $N$-body simulation represent that point distribution which is obtained when the Poisson sampling process discussed by Layzer (1956) and by Peebles (1980) is applied to the continuous density field. Indeed, the assumption of just such a relation between the discrete particles in the simulations and the underlying continuous density field is necessary to justify comparing statistical quantities such as the discrete particle-particle correlation function in these simulations with theoretical calculations of the evolution of the two-point corrrelation function of the continuous density field.

If the Poisson sampling process is an accurate model of the role of discreteness in the simulations, then the way in which the Press–Schechter mass functions are modified by the discreteness is calculable. Although, by assumption, all clumps have exactly the same overdensity ($\sim$178 times the background density), the calculation is made easier by as-suming that, to first order, all clumps have the same volume. For the initially scale free distributions studied by Efstathiou et al. (1988), this means that, to a good approximation, the Poisson sampled discrete Press–Schechter functions are given by

$$\frac{\eta(N)}{\sum_{N=1}^{\infty} N\eta(N)} = \int_0^{\infty} \frac{M^N e^{-M}}{N!} \, n(M) \, \mathrm{d}M, \quad (21)$$

where $\eta(N)$ is the probability that a Press–Schechter clump has $N$ associated particles, so that the denominator on the left is the average clump size. When $n = 0$ this integral is analytic:

$$\frac{\eta(N)}{\sum N\eta(N)} = \frac{\Gamma\left(N - \frac{1}{2}\right)}{\Gamma(N+1)\,\Gamma\left(\frac{1}{2}\right)} \frac{1}{\sqrt{1+M_*}} \left(\frac{M_*}{1+M_*}\right)^{N-1}, \quad (22)$$

where the sum is over all $N > 0$. Thus,

$$\frac{\sum_{N=1}^{\infty} \eta(N)}{\sum_{N=1}^{\infty} N\eta(N)} = \frac{2}{M_*} \left(\sqrt{1+M_*} - 1\right), \quad (23)$$

which has been simplified using the definition $(2n)!! = 2^n \, n!$, the identity $\Gamma(n+\frac{1}{2})/\Gamma(\frac{1}{2}) = (2n-1)!!/2^n$, and the Taylor series expansion of $(1-x)^{1/2}$. Requiring that $\eta(N)$ be normalized to unity implies that

$$\eta(N) = \frac{1}{2} \frac{\Gamma\left(N - \frac{1}{2}\right)}{\Gamma(N+1)\,\Gamma\left(\frac{1}{2}\right)} \left(\frac{M_*}{1+M_*}\right)^N \frac{\sqrt{1+M_*}}{\sqrt{1+M_*} - 1}. \quad (24)$$

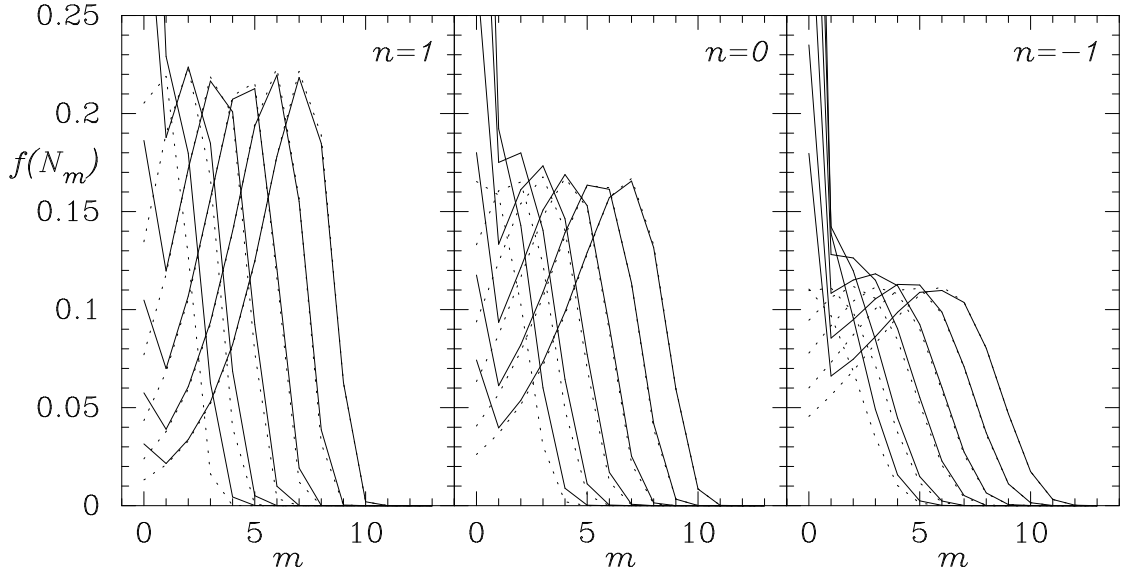

**Figure 3.** Multiplicity functions that are obtained after applying the Poisson sampling process of equation (21) to the Press–Schechter $n(M)$ distributions, for three choices of $n$, and a range of epochs. Plots show the fraction of particles at each epoch that belong to groups with $N_m$ members, where $2^{m-1} < N_m \leq 2^m$. Solid lines show the discrete clump multiplicity functions (obtained from eq. 21) for a range of expansion factors (all but the last one as shown in Fig. 8 of Efstathiou et al. 1988). Dotted lines show the corresponding curves for the continuous distribution $M^2 n(M) \log_e 2$.

For other values of $n$, equation (21) must be integrated numerically.

The panels in Fig. 3 show the functions that are obtained after applying the Poisson sampling process of equation (21) to the Press–Schechter $n(M)$ distributions, for a range of choices of $n$. The solid lines in each panel show the resulting discrete clump multiplicity functions (eq. 21) for a range of expansion factors (the first six of the last seven expansion times in the Efstathiou et al. simulations), and are chosen for comparison with Fig. 8 of Efstathiou et al. (1988). (As $M_*$ increases, the curves peak at larger values of $m$.) For comparison, the dotted lines show $M^2 n(M) \log_e 2$, where $n(M)$ is the continuous Press–Schechter distribution, at these same epochs.

For each $n$, the solid (discrete) and dotted (continuous) distributions have similar large clump tails, but differ increasingly as the clump size decreases. This difference decreases as the characteristic mass $M_*$ increases. This behaviour is easily understood since, as the simulation evolves, the characteristic mass increases, so the mass of a single particle as a fraction of the characteristic mass decreases. Thus, the effects of this lower mass cutoff, this discreteness, become less pronounced as the simulation evolves. The multiplicity functions in Fig. 3 are similar to those shown in Fig. 8 in Efstathiou et al. (1988). (Applying the Poisson sampling process to the modified mass functions given by Jedamzik 1995 increases the similarity between Fig. 3 here and the Efstathiou et al. figure significantly.) Thus, Fig. 3 suggests that the Poisson sampling process does, indeed, provide a good description of the relation between the particles in the $N$-body simulations and the underlying idealized continuous density field.

We have established that the Poisson sampling process of equation (21) provides a self-consistent way of modelling discreteness effects on the evolution of structure in the $N$-body simulations. So, we can use this sampling model to estimate the importance of discreteness on the distribution of pairwise velocities. This is done by summing over the discrete $\eta(N)$ distribution, rather than integrating over the continuous $n(M)$ distribution, when computing the analogue of equation (9). In the limit of vanishing separation, the velocity distributions obtained using the $\eta(N)$ distributions resemble those in Fig. 1 closely, with the agreement increasing at later times. This is expected since the continuous and the discrete multiplicity functions are quite different at early times (Fig. 3). For example, when $n = 0$, then the fractional difference between the mean square velocity in the continuous and the discrete cases is less than one per cent for expansion factors $a > 4$. The curves obtained using $\eta(N)$ have slightly more rounded peaks than those in Fig. 1. Since the discreteness provides a lower bound to the mass, this slight rounding of the peaks is consistent with the discussion at the end of Section 2.2. We conclude, therefore, that the effects of discreteness on the shapes of the pairwise velocity distributions in these simulations are not significant.

Measurements of $f(u;r)$ from observations of galaxy peculiar velocities (Peebles 1976) and $N$-body simulations of gravitational clustering (Fisher et al. 1994) show that, at least on highly nonlinear scales, an exponential distribution is a good approximation to the exact $f(u;r)$ distribution. This section suggests that the Press–Schechter approach is able to provide a simple, natural explanation for this agreement.

## 3  THE DISTRIBUTION OF PECULIAR VELOCITIES

This section considers the distribution $f(v)$ of peculiar velocities, which is the analogue of the Maxwell-Boltzmann velocity distribution that obtains in a perfect gas. It is, of course, different from the pairwise velocity distribution, $f(u)$, considered in the previous section.

### 3.1  The model

To begin, we summarize the relevant features of the model developed in the previous section. Press–Schechter clumps are assumed to be virialized singular isothermal spheres. The virialized assumption requires that the circular velocity, $V_c \propto M^{1/3}$. The isothermal assumption means that the circular velocity $V_c$ is independent of radius. Within a clump, the distribution of velocities relative to the center of mass is assumed to be Maxwellian. Whereas in the previous section clump-clump motions were irrelevant, they are the main source of uncertainty here. To compute $f(v)$ we need a model for these clump motions.

Consider the following simple model for these clump motions. Although it is not fully self-consistent, and a more detailed treatment, such as that developed by Bond & Myers (in preparation) may be more appropriate, this simple model will serve to provide some insight into the shape of the resulting velocity distribution. First, consider those clumps with exactly $N$ identical member particles. Let $c_j$ denote the velocity of the $j$th such clump, i.e., the motion of the center of mass of the $j$th clump. Let $s_{ij}$ denote the velocity, relative to the center of mass of the clump, of the $i$th particle in the $j$th clump. As in the previous section, assume that the distribution of $s$ is Maxwellian (with a dispersion that is related to $N$). In addition, assume that the distribution of $c$ is also Maxwellian, and that its dispersion is related to the dispersion of $s$ by some constant factor. Since $s$ and $c$ are both Maxwellian, $s + c$ is also Maxwellian (e.g. Binney & Tremaine 1987, p.485). Now allow for the Press–Schechter distribution of clump sizes $N$, and assume that the constant that relates the dispersion of $s$ to the dispersion of $c$ is independent of $N$. Finally, set the constant factor by requiring that the dispersion of $v \equiv s + c$, when averaged over all particles in all clumps, gives the three-dimensional velocity dispersion, $\sigma^2$.

As was the case in Section 2, the resulting distribution (now simply of peculiar velocities, $v = s + c$, not pairwise peculiar velocities, $u$) is given by convolving a Maxwellian with the distribution of circular velocities that is prescribed by the Press–Schechter approach. The resulting three dimensional velocity distribution is given by a relation like equation (5), with $u$ replaced by $v$, and where the Press–Schechter mass function is weighted by the number of particles, rather than by the number of pairs. That is, the probability $f(v)$ that a particle has velocity in the range $dv$ about $v$ is

$$f(v) = \int_0^\infty p(v|M)\, M\, n(M)\, \mathrm{d}M, \qquad (25)$$

where $p(v|M)$ is a Maxwellian with dispersion related to $M$, and $n(M)$ is the distribution of clump masses.

### 3.2  Comparison with $N$-body simulations

The distribution of peculiar velocities has not been studied as extensively as has the distribution of pairwise velocities. However, an extensive study, using $N$-body simulations, of clustering from an initially Poisson distribution has been made by Itoh, Inagaki & Saslaw (1988; 1993), and by Saslaw et al. (1990). In addition to measuring various properties of spatial clustering, they also measure the velocity distribution. Therefore, we will compare the model (eq. 25) with the peculiar velocity distribution they measure in their simulations. To do so, we must compute the Poisson Press–Schechter mass multiplicity function.

The Press–Schechter description of clustering from an initially Poisson distribution has been derived (Epstein 1983; Sheth 1995). The probability that a Poisson Press–Schechter has $N$-particles is given by the Borel distribution:

$$\eta(N) = \frac{(Nb)^{N-1} e^{-Nb}}{N!} \qquad \text{where} \quad b = \frac{1}{1+\delta_c}, \qquad (26)$$

and $\delta_c \approx 1.68/a$ is the usual Press–Schechter critical density that is necessary for collapse at the epoch $a$. This distribution of clump masses is in good agreement with the spatial distribution of particles in the Itoh et al. $N$-body simulations (Sheth 1995). At any epoch, the Poisson Press–Schechter model suggests that the spatial distribution of the particles is described by

$$P(N, \bar{n}V) = \frac{\bar{N}(1-b)}{N!} \left[\bar{N}(1-b) + Nb\right]^{N-1} e^{-\bar{N}(1-b)-Nb}, (27)$$

where $P(N, \bar{n}V)$ denotes the probability that a randomly placed cell of size $V$ contains exactly $N$ particles, $\bar{n}$ is the average number density of particles, $\bar{N} = \bar{n}V$ denotes the average number of particles in a $V$-sized cell, and $b$ increases monotonically from zero (initially) to some larger value (not greater than unity) as the clustering develops (Sheth 1995). Itoh et al. confirm the accuracy of equation (27) for the spatial counts in cells distribution. (Incidentally, Sheth, Mo & Saslaw 1994, and references therein, show that eq. 27 also describes the spatial distribution of optical and infrared galax-ies well.) Given that the spatial distribution of the particles is well described by the Press–Schechter approach, we can ask if the velocity distribution is also.

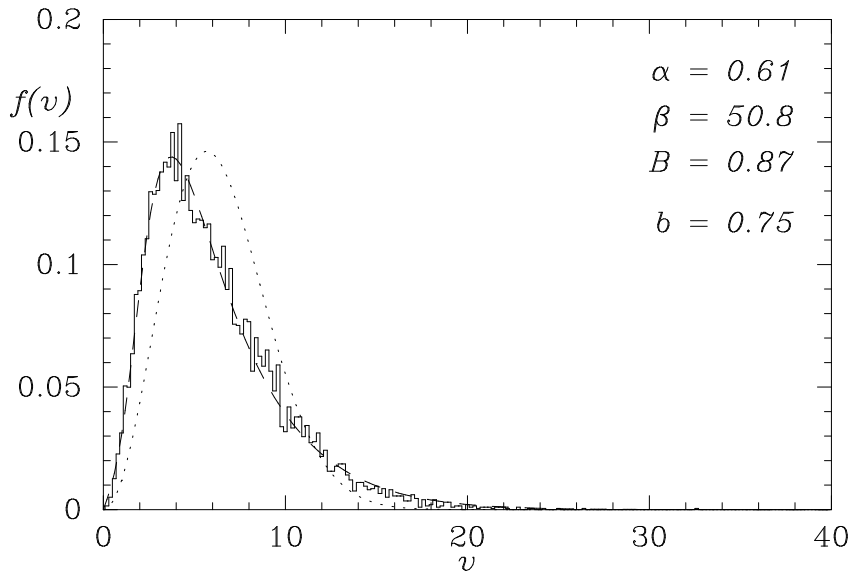

**Figure 4.** Example of the distribution function that corresponds to the Poisson Press–Schechter model. Histogram shows equation (25) when it is required to have the same velocity dispersion as the simulations when $10^4$ particles have collapsed into Press–Schechter clumps, in such a way that the spatial distribution of the particles is described by equation (27) with $b = 0.75$. Dotted line shows a Maxwellian that has the same dispersion as the histogram. Dashed line shows the best fit of equation (28) to the $N$-body histogram; top right corner shows the values of the parameters, $\alpha, \beta$, and $B$, that provide the best-fit to the histogram shown. These best-fit parameters are very similar to those for the $N$-body curve. Velocities are in "natural units" for ease of comparison with the Itoh et al. (1993) $N$-body simulations. This is why $\beta$, rather than $\sigma^2$, is shown.

Fig. 4 shows a comparison of the Press–Schechter model for $f(v)$ with the velocity distribution that is measured in the Itoh et al. $N$-body simulations of clustering from an initially Poisson distribution of identical particles that have no initial peculiar velocities. (These $10^4$-body simulations are described in more detail by Itoh et al. 1993). For ease of comparison with published results, all velocities are in "natural units" (Saslaw et al. 1990; Itoh et al. 1993). The histogram in Fig. 4 shows the Press–Schechter model (eq. 25) when it is required to have the same dispersion as the simulations, at that epoch when the $10^4$ particles have collapsed into Press–Schechter clumps, so that the spatial distribution of the particles is described by equation (27) with $b \approx 0.75$. (In the simulations, this occurs at an expansion factor $a/a_0 = 8$.) For comparison, the dotted line shows a Maxwellian that has the same dispersion as the histogram. The histogram is significantly different from the Maxwellian.

The $f(v)$ distribution that is measured in the simulations is represented by the dashed line, which shows the best fit of

$$f(v)\,dv = \frac{\alpha\sigma^2(1-B)}{\Gamma(\alpha v^2+1)} \left(\alpha\sigma^2(1-B)+\alpha v^2 B\right)^{\alpha v^2-1}$$
$$\times\ e^{-\alpha\sigma^2(1-B)-\alpha v^2 B}\ 2\alpha v\ dv \qquad (28)$$

to the simulation histogram. Equation (28) was first obtained by Saslaw et al. (1990), and has been shown to fit the velocity distributions in the Itoh et al. $N$-body simulations with remarkable accuracy (Inagaki, Itoh & Saslaw 1992; Itoh et al. 1993). In equation (28), $\alpha$ and $B$ are free parameters, and $\sigma^2$ is the three dimensional mean square velocity. The top right corner of the plot shows the values of the parameters that provide the best-fit to the Press–Schechter histogram shown. These best-fit values are very similar to (within a few percent of) those measured in the $N$-body simulations (Itoh et al. 1993). Similar agreement is obtained for all values of $b \lesssim 0.8$; at later times ($b \gtrsim 0.8$) it is not clear that these small ($10^4$-body) simulations remain statistically homogeneous. Since the best fitting values of equation (28) to the $N$-body $f(v)$ are so similar to those of the Press–Schechter model (eq. 25), and in both cases the fits are very accurate, we conclude that equation (25) provides a good fit to the velocity distribution in the simulations.

Since the Press–Schechter distribution evolves as the clustering develops, the model (eq. 25) can describe the evolution of the velocity distribution in the simulations. As the spatial distribution tends to the initial Poisson distribution (i.e., as $b \to 0$), then equation (25) reduces to a Maxwellian, as required by linear theory. At later times, the distribution becomes significantly different from Maxwellian. This evolution simply reflects the changing fraction of particles in small clumps (small velocities for free particles that lose energy to the expansion of the universe) relative to the fraction in massive virialized systems (having high virial velocities). Thus, as the clustering develops, the velocity distribution becomes skewed relative to the Maxwellian. This evolution is consistent with that measured in the simulations (Itoh et al. 1993).

Given the accuracy with which equation (28) fits equation (25), it is instructive to consider its functional form further. In effect, the parameter $\alpha$ simply rescales all velocities. Saslaw et al. (1990) hypothesize that the relation $\alpha'\sigma^2 = Gm\langle 1/r\rangle = (Gm/a)\langle 1/x\rangle$, where $m$ is the mass of each particle in their simulations, $\langle 1/x\rangle$ is some mea-sure of the average separation between particles measured in comoving coordinates, and $a$ denotes the expansion factor since some fiducial time (usually taken to be the initial time, $a_0 = 1$), should always be satisfied. Since $Gm\langle 1/x \rangle$ is some constant, their hypothesis means that $a\sigma^2 \equiv a'a\sigma^2$, should be constant. This hypothesis appears to be in good agreement with what Itoh et al. (1993) measure in their $N$-body simulations of clustering from cold Poisson initial conditions (see Model S in their Fig. 11; they use "natural units" in which their $\beta$ is proportional to our $\sigma^2$ and their $b_{\rm velocity}$ is our $B$). Typically, the agreement with their hypothesis is good after the simulations are a few ($\gtrsim 4$) expansion factors larger than the initial size.

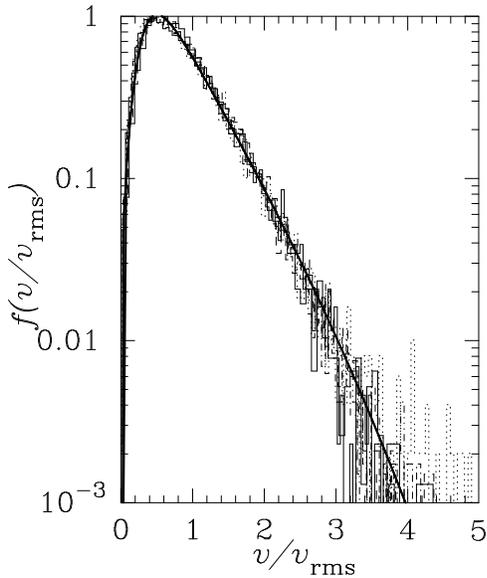

**Figure 5.** Approximately self-similar evolution of the velocity distribution in $N$-body simulations of clustering from an initially cold Poisson distribution. Curves show $f(v/v_{\rm rms})$ as a function of $v/v_{\rm rms}$ for a range of expansion factors. The bold solid line shows that equation (28), with $B = 0.87$, provides a good description of $f(v)$.

Consider those epochs over which $a\sigma^2$ is constant. If $B$ is also constant at these epochs, then equation (28) evolves self-similarly. Fig. 11 of Itoh et al. (1993) shows that indeed, at late times, $B$ is roughly independent of expansion factor. If we consider these Itoh et al. simulations (from Poisson initial conditions) as being roughly equivalent to simulations from Gaussian $n = 0$ initial conditions, then the Davis & Peebles scaling solutions should apply to these (initially Poisson) simulations also. In this respect, the self-similar behaviour (at times when $a\sigma^2$ and $B$ are both approximately constant) of equation (28) can be thought of as approximating the self-similar $f(v)$ distribution that must obtain in the Davis & Peebles description.

Fig. 5 shows that, to a good approximation, the evolution of $f(v)$ in these Itoh et al. simulations is self-similar. The histograms show $f(v/v_{\rm rms})$ as a function of $v/v_{\rm rms}$ when the radius of the simulation sphere is 4 (solid), 5.66 (dashed), 8 (dot-dashed), 11.3 (dotted), 16 (dot-dot-dashed), and 22.63 (solid) times the initial radius. The bold solid line that provides a good fit to all the histograms shows equation (28) with $B = 0.87$; at each epoch it has been rescaled to describe $f(v/v_{\rm rms})$.

So, equation (28) admits (at least approximately) self-similar behaviour, and this self-similar behaviour is in good agreement with the $N$-body simulations. However, as shown in the previous section, self-similar behaviour arises naturally in the Press–Schechter model developed in this paper. Thus, the close agreement between the shapes of equations (25) and (28) imply that the Press–Schechter model developed in this paper is able to explain the shape, and the self-similar evolution, of the distribution, $f(v)$, of peculiar velocities.

The close agreement between the distribution (eq. 28) obtained by Saslaw et al. (1990) and that considered here (eq. 25) is striking. However, there are two important differences between the Saslaw et al. (1990) derivation of $f(v)$ and that presented here. Saslaw et al. used the assumption that fluctuations in kinetic energy are proportional to fluctuations in potential energy to derive equation (28) from the spatial distribution function (eq. 27). In particular, they assumed that $\langle V^2 \rangle \propto \langle GM/r \rangle$, where the average is taken over randomly placed cells of some given size. They argued that this implies that $\langle V^2 \rangle \propto \langle M \rangle$. That is, they assumed that $\langle M/r \rangle$ was separable, that $\langle M/r \rangle = \langle M \rangle \langle 1/r \rangle$, where the average was taken over all space. However, our model of virialized, isothermal, Press–Schechter clumps, is constructed differently. In our model, $\langle V^2 \rangle \propto \langle M^{2/3} \rangle$, where the average is taken over the Press–Schechter distribution of clump sizes. In this case, we showed that the averages are not separable.

Secondly, since Saslaw et al. derive equation (28) for $f(v)$ from equation (27) for $f(N)$, their derivation suggests that $b = B$. In our model there is no reason for this equality to hold, and, indeed, the simulations show that $B > b$ always. In our model $B$ is simply a free parameter that is determined by the best-fit of equation (28) to the $N$-body or to the Press–Schechter $f(v)$ histogram. Furthermore, whereas $b$ increases as the simulation evolves, Fig. 5 suggests that, after initial transients (due to the fact that in the simulations, particles have no initial velocities) have died out, then, to a very good approximation, $B$ is constant, and the subsequent evolution of $f(v)$ is self-similar.

It is well known that equation (27) for $f(N)$ is affected by random sampling (e.g. Lahav & Saslaw 1992). On the other hand, equation (25) shows that $f(v)$ in our model is the same for a subsample drawn randomly from some parent distribution as it is for the parent distribution. The simulations show that, indeed, $f(v)$ is hardly changed by random dilution. In our model, therefore, the difference between $b$ [measured from $f(N)$] and $B$ [from $f(v)$] can be used to provide information about the true underlying distribution when only a randomly diluted subsample has been measured.

Although Fig. 4 suggests that equations (25) and (28) are quite similar, they behave differently in the large $v$ limit. These differences appear when $f(v) \lesssim 10^{-4}$ and so they are beyond the resolution of the $10^4$-particle simulations against which $f(v)$ has been tested. Given the nature of the derivation of equation (25), and its *ad hoc* treatment of the velocities between clumps in particular, we will not study the behaviour of this large $v$ tail further. We simply note that the model presented here serves to illustrate that the distribution of velocities in the $N$-body simulations can be un-derstood by making rather simple assumptions about the properties and motions of Press–Schechter clumps.

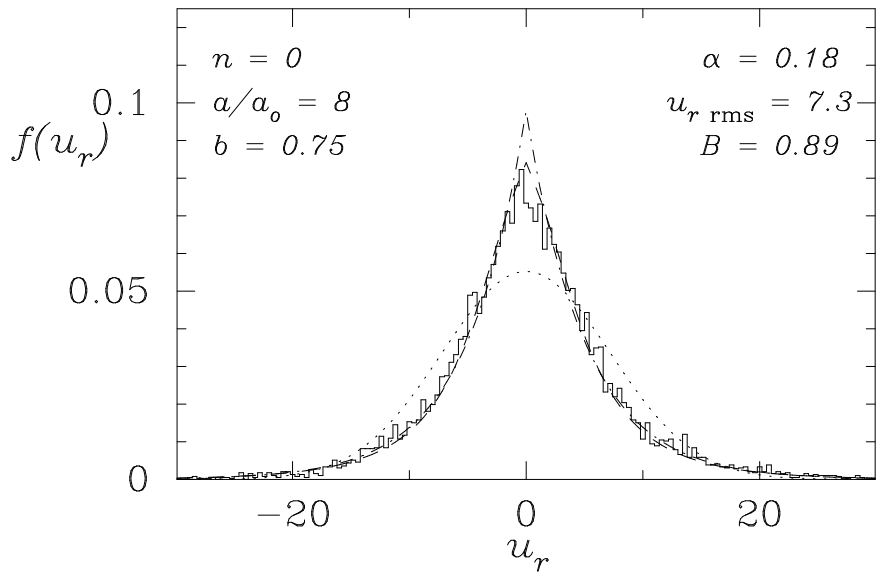

**Figure 6.** Example of the distribution function of pairwise relative peculiar velocities as a result of clustering from an initially Poisson distribution. Histogram shows this distribution when the expansion factor, $a/a_0$, in the $N$-body simulations described in the text, is 8. At that epoch, the spatial distribution of all particles is well fit by equation (27) with $b = 0.75$. Dot-dashed line shows an exponential distribution that has the same dispersion as the histogram. Dotted line shows a Gaussian that has this same dispersion. Dashed line shows the distribution that is obtained by integrating equation (28) over directions perpendicular to the line of sight. Top right shows the value of the rms pairwise peculiar velocity, $u_{r\,\rm rms}$, in "natural units" (Inagaki et al. 1992), as well as the values of the other parameters in equation (28), that are necessary to provide a good fit to the histogram.

It is worth noting that the particular prescription we have chosen to specify the distribution of clump-clump velocities is not completely *ad hoc*. Firstly, it is clear that the clumps must have some net velocities with respect to each other. This is easily verified by locating the Press–Schechter clumps in the simulations, and then computing the net velocity of the center of mass of each clump. The shape of the distribution of clump velocities so obtained is better described by equation (28) than by, for example, a Maxwellian. However, we have shown that equation (28) is a good approximation to the distribution that is obtained by adding up a number of Maxwellian distributions with a (Press–Schechter) distribution of dispersions. In effect, our model assumes just such a prescription for clump-clump motions. Since the motion of a clump of mass $M$ is assumed to be drawn from a Maxwellian with dispersion related to $M$, on average in this model, more massive clumps move faster than less massive clumps. It is intriguing that massive clumps have been measured to move faster than less massive clumps in simulations of a CDM (rather than an initially Poisson) universe (see caption to Fig. 5a in Zurek et al. 1994).

Finally, recall that the distribution for $f(v)$ that is defined by equation (28) is similar to that obtained from the Poisson Press–Schechter model developed in this section, for all measured values of $b$ (as an example, Fig. 4 shows this to be true when $b \approx 0.75$). However, notice that the prescription for obtaining $f(v)$ from the Press–Schechter approach is very similar to that for obtaining $f(u)$; both are integrals over Maxwellians having a Press–Schechter distribution of dispersions. The only difference between $f(u)$ and $f(v)$ is that, whereas the pairwise distributions are obtained by weighting the Press–Schechter mass functions by the number of pairs ($\propto M^{4/3}$ for truncated, singular isothermal spheres), the $f(v)$ distribution in this section weights by the number of particles ($\propto M$). If this difference in weighting terms is not significant, then, with the appropriate redefinitions of parameters, the distribution that is obtained by integrating equation (28) over directions perpendicular to the line of sight should also provide a good approximation to the Poisson Press–Schechter distribution of pairwise velocities, $f(u_{\rm r};r)$.

Fig. 6 shows an example of the line of sight pairwise velocity distribution that is measured in the Itoh et al. simulations. Since these simulations have only $10^4$ (identical) particles, Fig. 6 shows the distribution of line of sight, pairwise velocity differences for all pairs separated by less than some scale $r_{\rm cut}$. Although it includes contributions from pairs of varying separations, and so differs from the usual $f(u;r)$ which is usually computed as a function of pair separation, it should provide a reasonable description of the shape of the more conventional $f(u;r)$ on scales smaller than that of a typical clump. To insure that this is the case, Fig. 6 is constructed with $2(3/4\pi r_{\rm cut}^3) = 300\rho_{\rm b}$, since clumps are assumed to have an overdensity of at least $178\rho_{\rm b}$.

The histogram shows this $f(u;r)$ distribution when the simulation has expanded to $a/a_0 = 8$ times its initial radius. That is, Fig. 6 shows the pairwise velocity distribution $f(u;r)$ when the distribution of peculiar velocities themselves, $f(v)$, is given by Fig. 4. Dot-dashed and dotted lines show exponential and Gaussian distributions that have the same dispersion as the histogram. Whereas the exponential distribution provides a much better fit to the histogram than the Gaussian, it has a higher, sharper peak than the histogram. The reason for this sharper peak was discussed in the previous section.

The dashed line that provides an even better fit to the histogram is obtained by integrating equation (28) over directions perpendicular to the line of sight. The top right of Fig. 6 shows the value of the rms pairwise peculiar velocity, $u_{rrms}$, in "natural units" (Inagaki et al. 1992), as well as the values of the other parameters in equation (28), that are necessary to provide a good fit to the histogram at this epoch. That the dashed line in Fig. 6 fits the histogram so well suggests that the Press–Schechter model of virialized isothermal clumps, when used to describe clustering from an initially Poisson distribution is, substantially, correct.

Fig. 6 shows that the shape of the pairwise velocity distribution for a distribution of particles that has evolved from an initially Poisson distribution, like that for a density field that was initially a scale-free $n = 0$ Gaussian random field, is well approximated by an exponential. Insofar as clustering from an initially Poisson distribution resembles clustering from an $n = 0$ Gaussian density field, this Poisson Press–Schechter model provides another test of the importance of discreteness effects. This is because an initially Poisson distribution has a natural lowest mass scale—that of a single particle. Fig. 6 shows that discreteness effects may alter the small velocity part of the velocity distribution, whereas the large velocity tail is not strongly affected. These differences in the peak of the distribution are consistent with the discussion at the end of the previous section, regarding the effects of a small-mass cutoff in the mass distribution.

Having established that the Poisson Press–Schechter model provides a good description of the pairwise velocity distribution, and of the distribution of peculiar velocities themselves, in (initially Poisson) $N$-body simulations, we now consider the observations. Raychaudhury & Saslaw (1995) show that the observed peculiar velocity distribution function of a representative sample of galaxies (the Matthewson et al. 1991 spirals, and the Dressler et al. 1987 ellipticals) is well described by integrating equation (28) over directions perpendicular to the line of sight. The close agreement between equations (28) and (25) (Fig. 4), and the fact that integrating a functional form like equation (25) over directions perpendicular to the line of sight gives a distribution that is nearly exponential (dashed and dot-dashed curves in Fig. 6), shows that the $f(v_{\rm r})$ distribution used by Raychaudhury & Saslaw is well approximated by an exponential distribution. Thus, the Raychaudhury & Saslaw results can be used to show that the exponential distribution provides a good approximation to the observed $f(v_{\rm r})$ distribution. This accuracy of the exponential distribution has also been noted by Bahcall, Gramann & Cen (1994), in their study of the $f(v)$ distribution of galaxies. It is intriguing that, as with the spatial distribution of galaxies, the approximately exponential distribution of galaxy peculiar velocities is easily modelled by the Poisson Press–Schechter approach.

## 4 DISCUSSION

Section 2 used the Press–Schechter description of nonlinear clustering to obtain an expression for the distribution of pairwise relative peculiar velocities. The derivation assumed that Press–Schechter clumps are virialized, truncated, singular isothermal spheres, that all have the same average density. Section 2 argued that if the velocity distribution within each Press–Schechter clump is Maxwellian, then the pairwise velocity distribution (when viewed along the line of sight and averaged over many different clumps) can be obtained by convolving Gaussians having a range of dispersions. It showed that the distribution of dispersions can be related to the Press–Schechter distribution of clump masses. Fig.s 1 and 2 showed that this Press–Schechter model can explain the exponential shape of the line of sight distribution of pairwise velocities that is measured in $N$-body simulations of gravitational clustering. This exponential shape also provides a good description of the pairwise velocity distribution of galaxies.

Section 2 also showed that, on small scales, the pairwise velocity dispersion in this model increases with scale. Since the size of a Press–Schechter clump, $r_{\rm vir}$, is only proportional to the cube root of the mass, the increase of the dispersion of $f(u; r)$ with increasing pair separation, $r$, is expected to be small. The dispersions shown in Fig. 2 (corresponding to scales on which $\xi \sim 10$) are larger than those which obtain in the zero-separation limit (eq. 20). On these small scales, this scale dependence is in qualitative agreement with that suggested by the cosmic virial theorem (e.g. Peebles 1980; Efstathiou et al. 1988).

The $N$-body simulations show that as the separation between pairs increases and the pairwise velocity dispersion increases, the distribution $f(u; r)$ becomes slightly skew (Fig. 6 of Efstathiou et al. 1988). As the pair separation increases, the two members of any given pair are increasingly likely to be drawn from different clumps. These clumps need not be moving with the same net peculiar velocity, nor need they necessarily have comparable circular velocities. To date, there is no good model for clump-clump motions, so that the scale dependence of this skewness cannot be calculated explicitly from the model presented in this paper. However, a skew distribution results when several Gaussians with differing means and variances are added together. Thus, in our model, this skewness is easily explained. Moreover, it is possible to use the model presented here to constrain the clump-clump velocity distribution, though it has not been done here. This constraint follows from requiring that, in the limit of very large separations, when the two members of a given pair are almost certainly drawn from different clumps, then the line of sight pairwise velocity distribution must reduce to the form that is expected from models of bulk flows.

Section 3 considered a simple model for the motions of Press–Schechter clumps. Using this model, it provided a description of the distribution of peculiar velocities $f(v)$, rather than pairwise peculiar velocities $f(u)$. Fig. 4 showed that the model provides a good description of the evolution of the velocity distribution as an initially Poisson distribution clusters gravitationally. Like the Press–Schechter model for the line of sight $f(u)$, the line of sight $f(v)$ distribution is essentially a convolution of a Gaussian with the Press–Schechter distribution of dispersions, times a weighting term. The essential difference between the two distributions is in this weighting term; it is $\propto M$ (to account for the total number of particles) for $f(v)$, and $\propto M^{4/3}$ (for the total number of pairs in an isothermal clump) for $f(u)$.

Given that the essential difference between $f(u)$ and $f(v)$ is in the weighting term, the close agreement in Fig. 6 between the dot-dashed curve (exponential distribution) and the dashed curve (integral of eq. 28 over directions perpendicular to the line of sight) has an interesting implication. Fig. 4 showed that equations (25) and (28) are extremely similar. Thus, the dashed curve in Fig. 6 is equivalent to integrating equation (25) over directions perpendicular to the line of sight. In other words, the dashed curve is a good approximation to the distribution that is obtained by summing up Gaussian distributions with a Press–Schechter distribution of dispersions, where each Gaussian is weighted by the number of particles rather than by the number of pairs. However, weighting by the number of pairs gives a distribution that is similar to an exponential (Fig. 1; Section 2). So, the close agreement between the dashed and the exponential (dot-dashed) curves in Fig. 6 shows that the shape of the integral over the Press–Schechter distribution of dispersions is relatively insensitive to whether one is weighting by the number of pairs, or by the number of particles.

The Appendix showed that, for the special case when clumps have a 'tophat' density profile, then weighting by the number of pairs is the same as weighting by the number of particles. For other density profiles, these two weighting functions will differ. Thus, Fig. 6 shows that the exponential shape of the pairwise velocity distribution is primarily a result of the Press–Schechter distribution of dispersions, and is not so sensitive to the density profile within clumps. Alternatively, if the density profile is fixed, then one can group the weighting function with the Gaussian distribution of (line of sight) velocities within clumps. In this case, Fig. 6 can be interpretted as evidence that the exponential shape will result even if the line of sight velocity distribution within a given clump is not exactly Gaussian. In other words, the exponential shape of $f(u)$ is not so sensitive to the exact density profile within clumps, nor is it particularly sensitive to the exact velocity distribution within each clump. Provided that the distribution of velocity dispersions within clumps is given by something like the Press–Schechter distribution, the exponential shape will result so long as the density profiles of clumps are something between tophats and isothermal spheres, and the velocity distribution within each clump is approximately Maxwellian.

The distribution of dispersions is a consequence of assuming that at any epoch, the circular velocity of a Press–Schechter clump may be related to its mass: $V_c \propto M^{1/3}$. This relation appears to be in good agreement with that measured in $N$-body simulations (e.g. Bond & Myers, in preparation). However, the $V_c \propto M^{1/3}$ relation is not consistent with the usual stability assumption that, after virializing, clumps, on average, do not change their physical size as the Universe continues to expand (e.g. Davis & Peebles 1977; Efstathiou et al. 1988; Hamilton et al. 1991; Nityananda & Padmanabhan 1993). The reason for this inconsistency is clear. It arises from the fact that not all clumps of a given mass at a given epoch will have formed at that epoch; some may have formed at an earlier epoch.

As a particular example, consider two clumps, each having the same mass, $M$, at a specified epoch, say, the present. Consider the case when one of these clumps was formed (i.e., virialized at 178 times the background density) at the present epoch, but the other formed at some earlier epoch (i.e., since the time when it virialized, it has survived without merging into a larger clump). By hypothesis, the clump that was formed earlier will have virialized at 178 times the background density at the earlier epoch. However, the background density decreases as the Universe expands. Since the background density at the time of formation determines the size, and so the circular velocity of the clump, then if the clustering is stable, the clump that was formed earlier will have a higher circular velocity than the clump of the same mass, $M$, that was formed at the present epoch.

Thus, if the clustering is stable, then at any given epoch, there should be some scatter around the $V_c \propto M^{1/3}$ relation. Since $V_c \propto \rho_b^{1/6}$, it is clear that this scatter will not be large. Nevertheless, it may change the slope of the relation between the circular velocity and the mass. The discussion above suggests that as a result of this scatter, $V_c \propto M^\alpha$, where $\alpha \lesssim 1/3$, will be more accurate. Bond & Myers (in preparation) find that $\alpha \approx 0.29$. Since this scatter is related to the probability that a clump of mass $M$ at some earlier epoch is still of size $M$ at some later epoch, it can be calculated self-consistently from the Press–Schechter approach. Thus, in principle, the Press–Schechter merger probabilities (Bond et al. 1991; Lacey & Cole 1993) can be used to quantify the scatter around the $V_c \propto M^{1/3}$ relation. (In fact, the merger probabilities show that the probability of zero accretion during some interval d$t$ is zero. However, physically, one expects that a small amount of accretion in time d$t$ will not change the density of the clump significantly. It is this notion, that a 'small' amount of accretion will not affect the density of a given clump significantly, which introduces the scatter around the density equals $178\rho_b$ relation.)

Since this scatter is small (recall that $V_c \propto \rho_b^{1/6}$), while it may affect the scale dependence of the distribution of pairwise relative velocities, it is unlikely to affect the shape of this distribution. Therefore, despite the minor inconsistency between the stable clustering prescription and the Press–Schechter model for the circular velocities of clumps, the conclusion that the Press–Schechter model can easily explain the exponential shape of the distribution of pairwise relative velocities along the line of sight remains valid.

There is another minor discrepancy between the stable clustering hypothesis and the Press–Schechter model discussed above. It arises from the fact that we assumed that all Press–Schechter clumps were isothermal spheres, independent of the slope $n$ of the initial power spectrum. In fact, stable clustering and the scaling solutions of the BBGKY hierarchy (Davis & Peebles 1977; Pebbles 1980) require that, in the highly nonlinear regime, the density within a Press–Schechter clump should scale as $r^{-\gamma}$, with $\gamma$ a function of $n$, rather than as $r^{-2}$ as assumed in Section 2. However, as we have argued in this section, the exponential shape of the pairwise velocity distribution is relatively insensitive to the density profile within clumps. Thus, a more exact calculation that incorporates the $n$ dependence of the (small scale) density profiles within clumps will not change our conclusion that the Press–Schechter model can easily explain the exponential shape of $f(u)$.

Although this paper has focussed on self-similar models for the initial density field (and so on cosmological models that have critical density), equation (5) is quite general; it may be used with any Press–Schechter multiplicity function. So, in principle, it may be used to compute the pairwise relative velocity distribution for cosmological models in which the density parameter is less than unity. As reliable peculiar velocity measurements on small nonlinear scales become available, equation (5) may provide a way to estimate $\Omega$.

However, there is a complication when comparing the Press–Schechter models of clustering from arbitrary initially Gaussian density fields with the observations. It arises from two uncertainties. The first is simply that observations measure velocities of discrete particles (galaxies), and it is not clear how to relate the Press–Schechter distribution of pairwise velocities in a continuous density field to that measured for discrete particles. For this reason Section 2 used a Poisson sampling argument to show that the effects of discreteness could be calculated self-consistently for the Efstathiou et al. $N$-body simulations (Fig. 3). In addition, Section 3 studied the Press–Schechter description of clustering from an initially Poisson distribution—since an initially Poisson distribution has a natural lowest mass scale. It argued that, insofar as clustering from an initially Poisson distribution should resemble that from an initially Gaussian, scale-free, $n = 0$ model, the Poisson Press–Schechter model can be treated as a self-consistent way of imposing the small-mass cutoff. The similarity of the Poisson Press–Schechter model to the $n = 0$ Gaussian model, and the accuracy of the Poisson sampling model for models with other values of $n$ suggests that, at least for the particles in the simulations, the effects of discreteness can be understood. Except for the peaks of the distributions, the overall shapes of the pairwise velocity distributions are hardly effected by the discreteness.

The second difficulty in relating these Press–Schechter models to the pairwise velocity distribution of galaxies is the well known problem of biasing. Whereas the simulations trace the evolution of dark matter, it is not clear that the baryonic matter from which (luminous) galaxies are thought to form, necessarily has the same distribution as the dark matter. In particular, whereas the $N$-body simulations suggest that the Press–Schechter model of virialized isothermal spheres is an adequate description of the distribution of sizes of dark matter clumps, it is not clear that it can also describe the distribution of sizes of galaxy clusters, or the distribution of galaxies within a cluster. If the number of galaxies associated with a given dark matter clump is related to the mass of that clump, and if the distribution of galaxies within a cluster traces the distribution of the dark matter, then the Poisson sampling model provides a way to predict the velocity distribution of galaxies from the Press–Schechter description of the dark matter. If the biasing is more complicated, then it is not clear that the shape of the velocity distribution of the dark matter need have any similarity to that of the galaxies. Therefore, it is intriguing that the line of sight pairwise velocity distributions of the dark matter distributions studied in this paper are, like the observations presently available, very well fit by exponential distributions.

## ACKNOWLEDGMENTS

I thank M. Itoh and S. Inagaki for providing data from their $N$-body simulations, Marc Davis and the Berkeley Astronomy Department for their hospitality, and Saleem Zaroubi for prompting me to think about the distribution of pairwise velocities.

**APPENDIX A: THE DISTRIBUTION OF PAIRS**

This Appendix computes the number of pairs as a function of scale, $s$, within a sphere of mass $M$ and radius $R$. Let $N(s)\,ds$ denote the number of pairs within a sphere of radius $R$ and mass $M$ that have separations in the range $ds$ about $s$. Then

$$N(s) \propto \int_0^R r^2\,dr \int_0^\pi \sin\theta\,d\theta \int_0^{2\pi} d\phi\, \rho(r) \int_0^{2\pi} \int_0^{\theta'_{\max}} \rho(r')\,s^2\,\sin\theta'\,d\theta'\,d\phi', \tag{A1}$$

where $\rho(x)$ denotes the density at a distance $x$ from the center of the sphere,

$$r'^2 = s^2 + r^2 - 2sr\cos\theta', \tag{A2}$$

and

$$-\cos\theta'_{\max} = \begin{cases} -\pi & \text{when } r < R - s, \\ \frac{R^2 - s^2 - r^2}{2sr} & \text{when } r > R - s. \end{cases} \tag{A3}$$

Of course, $N(s) = 0$ for all $s > 2R$. Clearly, $N(s)$ depends on the density distribution within the sphere. For a sphere of uniform density, $\rho(x) = 3M/4\pi R^3$ is independent of position, so

$$\begin{aligned}
N(s) &\propto \left(\frac{3M}{4\pi R^3}\right)^2 \int_0^{R-s} \int_0^\pi \int_0^{2\pi} r^2\,dr\,\sin\theta\,d\theta\,d\phi \int_0^{2\pi} \int_0^\pi s^2\,\sin\theta'\,d\theta'\,d\phi' \; + \\
&\qquad \left(\frac{3M}{4\pi R^3}\right)^2 \int_{R-s}^R \int_0^\pi \int_0^{2\pi} r^2\,dr\,\sin\theta\,d\theta\,d\phi \int_0^{\theta'_{\max}} \int_0^{2\pi} s^2\,\sin\theta'\,d\theta'\,d\phi' \\
&\propto \left(\frac{M^2}{R^3}\right) s^2 \left(3 - \frac{9}{4}\frac{s}{R} + \frac{3}{16}\frac{s^3}{R^3}\right).
\end{aligned} \tag{A4}$$

The total number of pairs in the clump is

$$\frac{1}{2}\int_0^{2R} N(s)\,ds \propto \frac{1}{2}\int_0^{2R} \left(\frac{M^2}{R^3}\right)\left(3 - \frac{9}{4}\frac{s}{R} + \frac{3}{16}\frac{s^3}{R^3}\right) s^2\,ds \propto \frac{M^2}{2}, \tag{A5}$$

as expected. The factor of $1/2$ is included to avoid counting pairs twice, and the upper limit on the integration is set by the fact that $N(s) = 0$ for separations larger than the diameter of the clump. The useful result is that, to lowest order in $s/R$, the number of pairs scales as $M^2/R^3$. Notice that, for a uniform sphere, $M \propto R^3$. This means that the number of pairs scales as the number of particles $M$.

The calculation for a truncated singular isothermal sphere of mass $M$ and radius $R$ can be performed analogously. The density, $\rho$, within a singular isothermal sphere scales as $1/r^2$, where $r$ is the distance from the center of the sphere. So, let $\rho = A/r^2$. The constant $A$ is set by requiring that the integral of the density over the sphere equal the mass, $M$. Then

$$\begin{aligned}
N(s) &\propto \left(\frac{M}{4\pi R}\right)^2 \int_0^R dr \int_0^\pi \sin\theta\,d\theta \int_0^{2\pi} d\phi \int_0^{2\pi} \int_0^{\theta'_{\max}} \frac{s^2\,\sin\theta'\,d\theta'\,d\phi'}{s^2 + r^2 - 2sr\cos\theta'} \\
&\propto \left(\frac{M}{4\pi R}\right)^2 \int_0^{R-s} \int_0^\pi \int_0^{2\pi} dr\,\sin\theta\,d\theta\,d\phi \int_0^{2\pi} \int_0^\pi \frac{s^2\,\sin\theta'\,d\theta'\,d\phi'}{s^2 + r^2 - 2sr\cos\theta'} \; + \\
&\qquad \left(\frac{M}{4\pi R}\right)^2 \int_{R-s}^R \int_0^\pi \int_0^{2\pi} dr\,\sin\theta\,d\theta\,d\phi \int_0^{2\pi} \int_0^{\theta'_{\max}} \frac{s^2\,\sin\theta'\,d\theta'\,d\phi'}{s^2 + r^2 - 2sr\cos\theta'} \\
&\propto 4\pi^2 s \left(\frac{M}{4\pi R}\right)^2 \left[\int_0^{R-s} \frac{dr}{r} \ln\left(\frac{s+r}{s-r}\right)^2 + \int_{R-s}^R \frac{dr}{r} \ln\left(\frac{R}{s-r}\right)^2\right].
\end{aligned} \tag{A6}$$

This must be evaluated for two separate cases. When $s > R - s$, or equivalently, when $R > s > R/2$, then

$$\begin{aligned}
N(s) &\propto 4\pi^2 s \left(\frac{M}{4\pi R}\right)^2 \left[\int_0^{R-s} 2\,\frac{dr}{r}\ln\left(\frac{s+r}{s-r}\right) + \int_{R-s}^s 2\,\frac{dr}{r}\ln\left(\frac{R}{s-r}\right) + \int_s^R 2\,\frac{dr}{r}\ln\left(\frac{R}{r-s}\right)\right] \\
&\propto \frac{s}{4}\left(\frac{M}{R}\right)^2 \left[2\ln(R-s)\ln\left(\frac{s}{R}\right) + \frac{(\ln R)^2}{2} - \frac{(\ln s)^2}{2} + \right. \\
&\qquad\qquad\qquad \left. \sum_{k=1}^\infty \frac{(-1)^{k+1}}{k^2}\left(\frac{R-s}{s}\right)^k + \sum_{k=1}^\infty \frac{1}{k^2}\left(\frac{s}{R}\right)^k \right].
\end{aligned} \tag{A7}$$

When $s < R - s$, or, equivalently, when $s < R/2$, then

$$N(s) \propto 4\pi^2 s \left(\frac{M}{4\pi R}\right)^2 \left[\int_0^s 2\,\frac{dr}{r}\ln\left(\frac{s+r}{s-r}\right) \int_s^{R-s} 2\,\frac{dr}{r}\ln\left(\frac{s+r}{r-s}\right) + \int_{R-s}^R 2\,\frac{dr}{r}\ln\left(\frac{R}{r-s}\right)\right]$$

$$= \frac{s}{4}\left(\frac{M}{R}\right)^2 \left[\left(\ln\left(\frac{R}{R-s}\right)\right)^2 + \pi^2 - 4\sum_{k=1}^{\infty}\frac{1}{(2k-1)^2}\left(\frac{s}{R-s}\right)^{2k-1} - 2\sum_{k=1}^{\infty}\frac{1}{k^2}\left(\frac{s}{R}\right)^k + 2\sum_{k=1}^{\infty}\frac{1}{k^2}\left(\frac{s}{R-s}\right)^k\right]. \tag{A8}$$

In either case, to lowest order, $N(s) \propto M^2/R^2$.

The text considers Press–Schechter clumps, which are spheres that all have the same average density, so that $M \propto R^3$. This means that, to lowest order, the number of pairs scales as $M^2/R^3 \propto M$ for a uniform sphere, and as $M^2/R^2 \propto M^{4/3}$ for truncated singular isothermal spheres. This is the result used in the text.

In this paper we are primarily interested in the way in which the number of pairs of a given separation, $s$, depends on the mass, $M$, and radius, $R$, of a given spherical clump. However, the calculations in this Appendix are also useful for the following problems. Suppose we are interested in calculating the two-point correlation function for particles within a sphere. To do so, we need to be able to estimate the number of pairs as a function of scale, under the assumption that the particles are distributed uniformly at random throughout the sphere. If the sphere were of infinite radius, and if the density within the sphere was $\rho = 3M/4\pi R^3$, then the number of particles that are a distance $s$ from a given particle is proportional to $4\pi\rho s^2$, so that the total number of pairs that have separation $s$ is half of $\rho V \times 4\pi\rho s^2 = (3M/R^3)\, s^2$. However, if the sphere were of finite size, then the number of pairs as a function of scale is modified. The correct expression for $N(s)$ for the uniform sphere case is equation (A4). In the limit as $R \to \infty$, it reduces correctly to the infinite sphere case. For separations that are on the order of 10% of the size of the sphere, the correction to the infinite sphere case is on the order of 10%.

Next, suppose that we are at the center of a sphere of approximately uniform density, and again, we are interested in estimating the correlation function for the particles within the sphere. Suppose, however, that our efficiency at detecting partices, our selection function, is a function of the distance to the particle. For convenience, suppose that that the efficiency decreases as $1/r^2$. Then the calculation in this Appendix provides a way of estimating the number of pairs as a function of separation, assuming that the underlying distribution is uniformly random. In this case, $\rho(r)$ plays the role of the selection function.